\documentclass[amsmath,11pt]{article}

\date{\empty}

\begin{document}

\title{\bf Newtonian nonlinear hydrodynamics and magnetohydrodynamics}

\author{Nicolaos K. Spyrou and Christos G. Tsagas\\ {\small Section of Astrophysics, Astronomy and Mechanics, Department of Physics}\\ {\small Aristotle University of Thessaloniki, Thessaloniki 54124, Greece}}

\maketitle

\begin{abstract}
We use covariant methods to analyse the nonlinear evolution of self-gravitating, non-relativistic media. The formalism is first applied to imperfect fluids, aiming at the kinematic effects of viscosity, before extended to inhomogeneous magnetised environments. The nonlinear electrodynamic formulae are derived and successively applied to electrically resistive and to highly conductive fluids. By nature, the covariant equations isolate the magnetic effects on the kinematics and the dynamics of the medium, combining mathematical transparency and physical clarity. Employing the Newtonian analogue of the relativistic 1+3 covariant treatment, also facilitates the direct comparison with the earlier relativistic studies and helps to identify the differences in an unambiguous way. The purpose of this work is to set the framework and take a first step towards the detailed analytical study of complex nonlinear systems, like non-relativistic astrophysical plasmas and collapsing protogalactic clouds.
\end{abstract}

\section{Introduction}\label{sI}
%%%%%%%%%%%%%%%%%%%%%%%%%%%%%%%%
General relativity is believed to describe strong gravitational fields and also to determine the large-scale dynamics of our universe. Nevertheless, when the gravitational field is weak and on scales well inside the Hubble length, Newtonian gravity remains a very good approximation. The same is also true when dealing with low temperature (cold) plasmas, where the effects of special relativity are negligible. All these mean that Newtonian physics remains a very dependable mathematical tool for a variety of astrophysical and cosmological studies. In particular, the theory can offer very useful insights regarding the behaviour of complex nonlinear systems, like a collapsing protogalactic cloud for example. Moreover, despite the fundamental differences between Newtonian and relativistic fluid dynamics, the two theories still share many close parallels. These analogies become more prominent and clear when using relative-motion descriptions, such as those the relativistic 1+3 covariant formalism and its Newtonian counterpart are based upon. Here we will use the latter.

The covariant approach to fluid dynamics assumes the existence of a unique vector field that represents the average velocity of the matter at each point in space, or at each spacetime event in the case of a relativistic study. The formalism offers a Lagrangian description, where every kinematic and dynamic quantity is decomposed down to its irreducible parts; a splitting that combines mathematical compactness and clarity with physical transparency. The fluid kinematics, in particular, are monitored through a scalar, a vector and a tensor field that respectively describe the average volume evolution, the rotational behaviour and the shear deformation of any given fluid element. The evolution of these variables is determined by a set of three propagation equations, supplemented by an equal number of constraints. Once the full (nonlinear) expressions have been obtained, the covariant formulae can be applied to any physical environment by simply adjusting the symmetries.

In the present article, we review the covariant approach to Newtonian hydrodynamics and provide the complete set of the nonlinear propagation and constraint equations that describe a bound, self-gravitating medium. We first consider the case of a barotropic fluid and examine the kinematic implications of inhomogeneity. This means looking at the gravitational collapse, the shear anisotropy and the rotational behaviour of the fluid. Our results show that overdensities tend to enhance the collapse, while underdensities act against contraction -- or tend to accelerate the expansion. We also find that, under the barotropic-fluid assumption, vorticity cannot be generated. At each step, we compare our Newtonian expressions to their relativistic counterparts and establish the main analogies and differences between the two. Exploiting the advantages of the covariant expressions, we apply our nonlinear formulae to the case of an imperfect medium, in an attempt to investigate the role of viscosity. Among others, we find that a viscous fluid will generally act as a source of rotation. Also, by involving the internal properties of the fluid, we discuss how hydrodynamic flows can be represented as purely gravitational motions and outline the potential applications of this dynamical correspondence.

With the full hydrodynamic equations in hand, we proceed to incorporate magnetic fields into our study. Introducing an electron-ion system and assuming overall charge neutrality, we derive the covariant magnetohydrodynamic (MHD) formulae for an electrically resistive fluid. These include, for the first time in the Newtonian limit, the covariant form of Maxwell's equations and allow for a direct comparison with their relativistic analogues. The evolution of the $B$-field is monitored by looking at both the isotropic and the anisotropic components of the magnetic pressure. Confining to a barotropic medium, we consider the effects of the field on the fluid kinematics. The magnetic implications for gravitational collapse, for example, are encoded in Raychaudhuri's equation. The latter reveals how increases in the pressure of the $B$-field assist the contraction by adding to the gravitational attraction of the matter. We also identify in covariant terms what is commonly referred to as ``magnetic braking'', and show how the effect results from the elasticity (i.e.~the tension) of the magnetic forcelines. As with hydrodynamics, we take every opportunity to compare our Newtonian expressions to their general relativistic partners and identify all parallels and differences between the two sets. Thus, in contrast with general relativity, we find that the magnetic pressure has no effect on Newtonian vorticity. In agreement with the relativistic analysis, on the other hand, the magnetic tension is found to affect rotation and act as a source of it. Finally, by assuming a perfectly conductive medium, we apply our results to the ideal MHD case, establish the pattern of the magnetic evolution in such an environment and also discuss how the magnetohydrodynamic equations can be reduced to pure hydrodynamic ones.

The main aim of this work is to introduce the key features of a formalism that will be subsequently used in nonlinear Newtonian hydrodynamic and magnetohydrodynamic studies. For this reason we have gone beyond the perfect-fluid approximation and incorporated viscosity effects into our equations. Similarly, the MHD formalism has been extended to allow for media of finite (nonzero) electrical resistivity. Our targets are nonlinear systems that are adequately described by the Newtonian theory. These include non-relativistic astrophysical plasmas and protogalactic clouds (of subhorizon size) that have decoupled from the background expansion and started to collapse.

\section{Covariant hydrodynamics}\label{sCHDs}
%%%%%%%%%%%%%%%%%%%%%%%%%%%%%%%%%%%%%%%%%%%%%%
The covariant approach to fluid dynamics dates back to the 1950s and the work of Heckmann, Sch\"{u}cking and Raychaudhuri~\cite{HS1}. The formalism was originally applied within the Newtonian framework before extended to general relativistic hydrodynamics and magnetohydrodynamics (see~\cite{EvE,TCM} for recent review articles and further references). In the present section we first review and later (see \S~\ref{ssPFs}-\ref{ssHFPGMs}) extend parts of~\cite{E1}, where the reader is referred for more details. Relative to that article, there are also several notational differences, reflecting the presentation changes that have taken place since the early 1970s. For an alternative, 4-dimensional covariant approach to Newtonian hydrodynamics we refer the reader to~\cite{CC}.

\subsection{Self-gravitating fluids}\label{ssS-GFs}
%%%%%%%%%%%%%%%%%%%%%%%%%%%%%%%%%%%%%%%%%%%%%%%%%%%
We use fixed space coordinates $\{x^a,\;a=1,2,3\}$ to define the metric tensor $h_{ab}$ of the Euclidean space, so that $v^2=h_{ab}v^av^b$ for any vector $v^a$. The above given metric and its inverse $h^{ab}$ -- with $h_{ac}h^{cb}=\delta_a{}^b$ and $\delta_a{}^a=3$, where $\delta_{ab}$ is the Kronecker symbol -- are used to raise and lower the tensor indices. When one uses Cartesian coordinates, as we will be doing here, $h_{ab}= \delta_{ab}$. Then, covariant and contravariant components coincide and partial derivatives are the `correct' spatial derivatives (see~\cite{E1,E2} for further details).\footnote{In a general frame $h_{ab}\neq\delta_{ab}$ and the covariant and contravariant tensor components do not always coincide. Then we need covariant, instead of partial, derivatives to compensate for the ``curvature'' of the system~\cite{E1,E2}.}

We adopt the fluid description, assuming the existence of a unique vector field representing the average velocity of the matter at each point. The 3-velocity field $v_a$ is tangent to the flow lines of the comoving (fundamental) observers. The time derivative of a tensorial quantity $T$ is given by the convective derivative $\dot{T}=\partial_tT+ v^a\partial_aT$, where $\partial_a=\partial/\partial x^a$. Thus, the convective derivative of the fluid velocity is
\begin{equation}
\dot{v}_a= \partial_tv_a+ v^b\partial_bv_a\,,  \label{dotva1}
\end{equation}
with $\partial_bv_a$ describing the spatial variations of the velocity field (e.g.~see~\cite{C}). Note that we adopt the Einstein summation convection, according to which repeated indices are summed. Like any second-rank tensor, the spatial derivative of $v_a$ decomposes as
\begin{equation}
\partial_bv_a= {1\over3}\,\Theta\delta_{ab}+ \sigma_{ab}+ \omega_{ab}\,,  \label{pbva}
\end{equation}
where $\Theta=\partial^av_a$, $\sigma_{ab}=\partial_{\langle b} v_{a\rangle}$ and $\omega_{ab}=\partial_{[b}v_{a]}$.\footnote{Round brackets in the indices denote symmetrisation, square indicate antisymmetrisation and angled ones define the symmetric and trace-free part of second-rank tensors. Therefore, $\partial_{\langle b}v_{a\rangle}= \partial_{(b}v_{a)}- (\partial^cv_c/3)\delta_{ab}$.} The tensor $\partial_bv_a$ monitors the relative motion between two neighbouriong fluid flow-lines.\footnote{The relative velocity vector ($\dot{x}^a$), between two neighbouring flow lines, is related to their connecting vector ($x^a$ -- connecting the same two particles at all times) via the transformation $\dot{x}^a=x^b\partial_bv^a$ (see~\cite{E1} for details).\label{rvv}} In particular, $\Theta$ determines the volume expansion, $\sigma_{ab}$ the shear deformation and $\omega_{ab}$ the rotational behaviour of a given fluid element. Positive values for $\Theta$ correspond to an expanding fluid, while negative ones indicate contraction. The volume scalar can also be used to define a representative length scale ($a$) along the flow lines by means of $\dot{a}/a= \Theta/3$. In cosmological studies, the aforementioned length scale corresponds to the scale factor of the universe. The antisymmetry of the vorticity tensor implies that we can define a vorticity vector by means of $\omega_a= \varepsilon_{abc}\omega^{bc}/2$, with $\varepsilon_{abc}$ representing the alternating tensor of the Euclidean space.\footnote{By construction the volume element (the Levi-Civita tensor) has $\varepsilon_{abc}=\varepsilon_{[abc]}$, with $\varepsilon_{123}=1$. Also, $\varepsilon_{abc}\varepsilon^{dqp}= 3!\delta_{[a}{}^d\delta_b{}^q\delta_{c]}{}^p$, which ensures that $\varepsilon_{abc}\varepsilon^{dqc}= 2!\delta_{[a}{}^d\delta_{b]}{}^q$, $\varepsilon_{abc}\varepsilon^{dbc}=2\delta_a{}^d$ and $\varepsilon_{abc}\varepsilon^{abc}=6$.} By construction $\omega_{ab}=\varepsilon_{abc}\omega^c$, ensuring that $\omega_{ab}\omega^b=0$. The vorticity vector also determines the rotation axis of the matter, namely the only direction that remains unaffected by the rotational motion~\cite{E1}. Finally, the shear and vorticity magnitudes are defined by $\sigma^2=\sigma_{ab}\sigma^{ab}/2$ and $\omega^2= \omega_{ab}\omega^{ab}/2=\omega_a\omega^a$ respectively~\cite{EvE}.

Assuming that $\Phi$ is the Newtonian gravitational potential, we use the velocity of the fluid to define the vector
\begin{equation}
A_a= \dot{v}_a+ \partial_a\Phi\,,  \label{Aa}
\end{equation}
which describes the combined action of gravitational and inertial forces. The vector $A_a$ corresponds precisely to the relativistic 4-acceleration and vanishes when the matter moves under inertial and gravitational forces alone~\cite{E1,E2}. The gravitational field is determined through a Poisson-like equation of the form
\begin{equation}
\partial^2\Phi= {1\over2}\,\kappa\rho- \Lambda\,,
\label{P}
\end{equation}
where $\partial^2=\partial^a\partial_a$ is the Laplacian operator, $\kappa=8\pi G$ represents the gravitational constant, $\rho$ is the density of the matter and we have allowed for a nonzero cosmological constant $\Lambda$ (in units of inverse-time squared).

\subsection{Nonlinear hydrodynamics}\label{ssNHD}
%%%%%%%%%%%%%%%%%%%%%%%%%%%%%%%%%%%%%%%%%%%%%%%%%
Using the convective derivative operator, decomposition (\ref{pbva}) and definition (\ref{Aa}), the nonlinear continuity equation and the Navier-Stokes formula associated with a self-gravitating fluid assume the covariant forms
\begin{equation}
\dot{\rho}=-\Theta\rho \hspace{15mm} {\rm and} \hspace{15mm}
\rho A_a= -\partial_ap- \partial^b\pi_{ab}\,,  \label{CEs}
\end{equation}
respectively~\cite{E1}. Note that $p$ is the isotropic and $\pi_{ab}$ is the anisotropic pressure of the medium (with $\pi_{ab}=\pi_{\langle ab\rangle}$). To close the system one requires the equations of state for the matter. These usually take the simple barotropic form adopted in \S~\ref{ssPFs}, or the phenomenological shape of Eq.~(\ref{pi}) in \S~\ref{ssIFs}, though in general they depend on additional thermodynamic variables. We also need a set of nonlinear formulae to describe the fluid kinematics. These comprise two sets of three propagation and constraint equations, which (like their relativistic counterparts) are obtained by applying the Newtonian analogues of the Ricci identities to the velocity vector of the fluid, namely by means of
\begin{equation}
\partial_{[t}\partial_{b]}v_a=0 \hspace{15mm} {\rm and} \hspace{15mm} \partial_{[c}\partial_{b]}v_a=0\,.  \label{NRi}
\end{equation}
The former of these integrability conditions leads to the propagation formulae. To be precise, the gradient of (\ref{Aa}) together with definition (\ref{dotva1}) and decomposition (\ref{pbva}), gives
\begin{equation}
\left(\partial_bv_a\right)^{\cdot}= -{1\over9}\,\Theta^2\delta_{ab}- {2\over3}\,\Theta(\sigma_{ab}+\omega_{ab})- \partial_b\partial_a\Phi+ \partial_bA_a- \sigma_{ca}\sigma^c{}_b+ \omega_{ca}\omega^c{}_b- 2\sigma_{c[a}\omega^c{}_{b]}\,.  \label{pbvadot}
\end{equation}
This expression contains collective information about the kinematical behaviour of the fluid. We decode this information by isolating the trace, the symmetric trace-free and the antisymmetric components of (\ref{pbvadot}).

We begin with the trace of Eq.~(\ref{pbvadot}), which by means of (\ref{P}) leads to the Newtonian version of the familiar Raychaudhuri equation,
\begin{equation}
\dot{\Theta}= -{1\over3}\,\Theta^2- {1\over2}\,\kappa\rho+ \partial^aA_a- 2(\sigma^2-\omega^2)+ \Lambda\,,  \label{Ray}
\end{equation}
that determines the expansion (or contraction) rate of the fluid.  Comparing the above to its relativistic counterpart (e.g.~see Eq.~(1.3.3) in~\cite{TCM}), we notice that only the density of the matter contributes to the gravitational mass and also note the absence of an $A_aA^a$-term in the right-hand side of (\ref{Ray}).

In an analogous way, the symmetric and trace-free component of (\ref{pbvadot}) provides the evolution formula of the shear
\begin{equation}
\dot{\sigma}_{ab}= -{2\over3}\,\Theta\sigma_{ab}- E_{ab}+ \partial_{\langle a}A_{b\rangle}- \sigma_{c\langle a}\sigma^c{}_{b\rangle}+ \omega_{c\langle a} \omega^c{}_{b\rangle}\,.  \label{sigmadot}
\end{equation}
Here, $E_{ab}=\partial_{\langle b}\partial_{a\rangle}\Phi$ represents the tidal part of the gravitational field and corresponds to the electric Weyl component of the relativistic treatment (compare the above to expression (1.3.4) in~\cite{TCM}).\footnote{The tidal field can be associated with a component of the gravitational potential that does not directly relate to matter and satisfies the Laplace equation (e.g.~see~\cite{BG}). Also note that there is no Newtonian analogue to the magnetic Weyl tensor, which reflects the absence of gravitational waves within the limits of Newton's theory.} The vorticity term, on the other hand, carries the distorting effect of the centrifugal forces~\cite{E1}. Also note that, in contrast with the relativistic analysis, there are no $A_{\langle a}A_{b\rangle}$ and $\pi_{ab}$ terms in the right-hand side of (\ref{sigmadot}).

We close the set of the propagation formulae with the skew part of (\ref{pbvadot}). The latter governs the rotational behaviour of the fluid element, either in terms of $\omega_{ab}$
\begin{equation}
\dot{\omega}_{ab}= -{2\over3}\,\Theta\omega_{ab}+ \partial_{[b}A_{a]}- 2\sigma_{c[a}\omega^c{}_{b]}\,,  \label{omegadot1}
\end{equation}
or in terms of $\omega_a$
\begin{equation}
\dot{\omega}_a= -{2\over3}\,\Theta\omega_a- {1\over2}\,{\rm curl}A_a+ \sigma_{ab}\omega^b\,,  \label{omegadot2}
\end{equation}
since $\omega_{ab}=\varepsilon_{abc}\omega^c$ by definition and ${\rm curl}v_a=\varepsilon_{abc}\partial^bv^c$ for any vector $v_a$. Therefore, when only gravitational/inertial forces are present and in the absence of shear, expressions (\ref{omegadot1}), (\ref{omegadot2}) imply  $\omega_a\propto a^{-2}$ and consequently constant angular momentum. Note that both of the above have the form of their relativistic counterparts (e.g.~compare (\ref{omegadot2}) to Eq.~(1.3.5) in~\cite{TCM}).

Expressions (\ref{Ray})-(\ref{omegadot2}) monitor the nonlinear evolution of the irreducible kinematical quantities of a Newtonian self-gravitating fluid in fully covariant terms. For a complete kinematical description, we need to supplement this set by an equal number of constraints. These come after contracting identity (\ref{NRi}b) with the permutation tensor of the space. Employing decomposition (\ref{pbva}), the result reads
\begin{equation}
\varepsilon_{cda}\partial^c\sigma_b{}^d+ \partial_b\omega_a- (\partial^c\omega_c)\delta_{ab}- {1\over3}\ \varepsilon_{abc}\partial^c\Theta= 0\,.  \label{kcons}
\end{equation}
The trace of the above, combined with the total antisymmetry of $\varepsilon_{abc}$, immediately leads to the familiar vorticity constraint
\begin{equation}
\partial^a\omega_a= 0\,,  \label{kcon1}
\end{equation}
guaranteeing that $\omega_a$ is a solenoidal vector. On the other hand, taking the symmetric and trace-free component of Eq.~(\ref{kcons}) we arrive at
\begin{equation}
{\rm curl}\sigma_{ab}+ \partial_{\langle b}\omega_{a\rangle}= 0\,,  \label{kcon2}
\end{equation}
where ${\rm curl}T_{ab}\equiv\varepsilon_{cd\langle a}\partial^cT_{b\rangle}{}^d$ for every symmetric and trace-free tensor of rank two. Finally, the antisymmetric part of (\ref{kcons}) leads to
\begin{equation}
{2\over3}\,\partial_a\Theta- \partial^b\sigma_{ab}+ {\rm curl}\omega_a= 0\,.  \label{kcon3}
\end{equation}
and provides a relation between the gradients of the three kinematic variables. The reader is referred to Eqs.~(1.3.6)-(1.3.8) in~\cite{TCM} for a comparison between the Newtonian and the relativistic kinematic constraints. Here, we simply note that in relativity $\omega_a$ is not generally a solenoidal vector. Further discussion on Newtonian covariant hydrodynamics can be found in~\cite{E1}.

So far our analysis applies to all situations where the fluid description is valid. Typical cosmological models, for example, have $\Theta>0$ and $\omega$, $\sigma$, $p\simeq0$. A non-rotating star, on the other hand, is characterised by $\omega\neq0$ and by $\Theta$, $\sigma\simeq0$, while $p\propto\rho^{\gamma}$ (with $\gamma=\,$constant) is a commonly used equation of state for the matter. Variations of the latter are also used in galactic studies, where observations indicate $\Theta\simeq0$ and we can use Oort's constants to estimate the associated shear and vorticity.

\subsection{Perfect fluids}\label{ssPFs}
%%%%%%%%%%%%%%%%%%%%%%%%%%%%%%%%%%%%%%%%
When the gravitational field is specified and an equation of state for the fluid has been introduced, expressions (\ref{CEs}), (\ref{Ray})-(\ref{omegadot2}) and (\ref{kcon1})-(\ref{kcon3}) provide the fully nonlinear covariant equations that monitor the hydrodynamic behaviour of a self-gravitating Newtonian fluid. Here, we will consider the case of a barotropic perfect fluid with $p=p(\rho)$ and $\pi_{ab}=0$. Under these conditions the Navier-Stokes equation reduces to
\begin{equation}
A_a= -{{\rm c}_s^2\over a}\,\Delta_a\,,  \label{bNS}
\end{equation}
where ${\rm c}_s^2\equiv{\rm d}p/{\rm d}\rho$ is the square of the adiabatic sound speed and $\Delta_a=(a/\rho)\partial_a\rho$. The latter is a dimensionless quantity that describes spatial variations (inhomogeneities) in the density of the fluid, as measured between two neighbouring flow lines (e.g.~see~\cite{E2}). Assuming, for simplicity, that both the sound speed and the scale factor have zero spatial dependence, the above leads to
\begin{equation}
\partial_bA_a= -{{\rm c}_s^2\over a^2}\,\Delta_{ab}\,,  \label{bpbAa}
\end{equation}
with $\Delta_{ab}=a\partial_b\Delta_a$. This variable is also dimensionless and, in contrast with the relativistic case, has zero skew part (i.e.~$\Delta_{[ab]}=0$). Thus, within the Newtonian framework, $\Delta_{ab}$ can be used to describe density perturbations (by means of the scalar $\Delta=\Delta_a{}^a= a\partial^a\Delta_a$) and shape distortions (via the symmetric and trace-free tensor $\Delta_{\langle ab\rangle}= a\partial_{\langle b}\Delta_{a\rangle}$) but no vortex-like (i.e.~vector) inhomogeneities. On using result (\ref{bpbAa}), expressions (\ref{Ray})-(\ref{omegadot1}) take the form
\begin{equation}
\dot{\Theta}= -{1\over3}\,\Theta^2- {1\over2}\,\kappa\rho- {{\rm c}_s^2\over a^2}\,\Delta- 2(\sigma^2-\omega^2)+ \Lambda\,,  \label{bRay}
\end{equation}
\begin{equation}
\dot{\sigma}_{ab}= -{2\over3}\,\Theta\sigma_{ab}- E_{ab}- {{\rm c}_s^2\over a^2}\,\Delta_{\langle ab\rangle}- \sigma_{c\langle a}\sigma^c{}_{b\rangle}+ \omega_{c\langle a} \omega^c{}_{b\rangle}  \label{bsigmadot}
\end{equation}
and
\begin{equation}
\dot{\omega}_{ab}= -{2\over3}\,\Theta\omega_{ab}- 2\sigma_{c[a}\omega^c{}_{b]}\,,  \label{bomegadot1}
\end{equation}
respectively. According to (\ref{bRay}), overdensities (i.e.~perturbations with $\Delta>0$) tend to enhance the gravitational collapse of the fluid, while underdensities support against it. In addition, following (\ref{bsigmadot}) and (\ref{bomegadot1}), the barotropic fluid can act as a source of shear anisotropy but does not generate vorticity. The same behaviour has also been seen in the relativistic studies (e.g.~see~\cite{TCM}). Here, the main difference is that rotation remains unaffected by the fluid pressure (compare expression (\ref{bomegadot1}) to Eq.~(3.2.8) in~\cite{TCM}). As a result of this, which is due to the zero curvature of the Euclidean space, vorticity can never grow in expanding Newtonian models with vanishing shear.\footnote{In general relativity, the rotational behaviour of the fluid also depends on its pressure. In particular, vorticity grows when the (dimensionless) adiabatic sound speed is greater than $\sqrt{2/3}$ (see~\cite{B} and also~\cite{TCM}).}

Finally, we note that one may monitor the acceleration or deceleration of an expanding Newtonian (barotropic) fluid by recasting Eq.~(\ref{bRay}) into the form
\begin{equation}
{1\over3}\,\Theta^2q= {1\over2}\,\kappa\rho+ {{\rm c}_s^2\over a^2}\,\Delta+ 2(\sigma^2-\omega^2)- \Lambda\,,  \label{q}
\end{equation}
where $q=-a\ddot{a}/\dot{a}^2$ is the deceleration parameter. The above also shows how ``voids'', namely underdense regions with $\Delta<0$, tend to accelerate the expansion by acting together with the vorticity and the (positive) cosmological constant.

\subsection{Imperfect fluids}\label{ssIFs}
%%%%%%%%%%%%%%%%%%%%%%%%%%%%%%%%%%%%%%%%%%
One may look at the implications of fluid viscosity by considering an imperfect medium with nonzero anisotropic pressure. Maintaining the $p=p(\rho)$ assumption of the previous section for simplicity, we introduce the phenomenological expression
\begin{equation}
\pi_{ab}= -\lambda\sigma_{ab}\,,  \label{pi}
\end{equation}
with $\lambda=\lambda(\rho,p)\geq0$ being the viscosity coefficient (e.g~see~\cite{E1}). When the latter is a slowly varying function, the above combines with constraint (\ref{kcon3}) to recast the momentum conservation law (see Eq.~(\ref{CEs}b)) into
\begin{equation}
A_a= -{{\rm c}_s^2\over a}\,\Delta_a+ {\lambda\over\rho} \left({2\over3a}\,\mathcal{Z}_a+{\rm curl}\omega_a\right)\,,  \label{iNS}
\end{equation}
where $\mathcal{Z}_a=a\partial_a\Theta$ describes inhomogeneities in the volume expansion/contraction. Thus, by exploiting the advantages of the covariant expressions (in particular by involving constraint (\ref{kcon3})), we were able to recast the viscosity term of (\ref{CEs}b) into a kinematical one. Proceeding as with the perfect fluid, we assume that both the sound speed and the scale factor depend solely on time. This allows the direct comparison of the two cases and leads to
\begin{equation}
\partial_bA_a= -{{\rm c}_s^2\over a^2}\,\Delta_{ab}+ {2\lambda\over3a^2\rho} \left(\mathcal{Z}_{ab}-\mathcal{Z}_a\Delta_b\right)+ {\lambda\over\rho}\left(\partial_b{\rm curl}\omega_a-{1\over a}\,\Delta_b{\rm curl}\omega_a\right)\,,  \label{ipbAa}
\end{equation}
with $\mathcal{Z}_{ab}=a\partial_b\mathcal{Z}_a$. Substituting the trace, the symmetric trace-free part and the skew component of the above into Eqs.~(\ref{Ray})-(\ref{omegadot1}), we arrive at
\begin{eqnarray}
\dot{\Theta}&=& -{1\over3}\,\Theta^2- {1\over2}\,\kappa\rho- {{\rm c}_s^2\over a^2}\,\Delta+ {2\lambda\over3a^2\rho}\left[\mathcal{Z}- \left(\mathcal{Z}_a+{3a\over2}\,{\rm curl}\omega_a\right) \Delta^a\right] \nonumber\\ &&-2(\sigma^2-\omega^2)+ \Lambda\,,  \label{iRay}
\end{eqnarray}
\begin{eqnarray}
\dot{\sigma}_{ab}&=& -{2\over3}\,\Theta\sigma_{ab}- E_{ab}- {{\rm c}_s^2\over a^2}\,\Delta_{\langle ab\rangle}+ {2\lambda\over3a^2\rho}\left(\mathcal{Z}_{\langle ab\rangle}-\mathcal{Z}_{\langle a}\Delta_{b\rangle}\right) \nonumber\\ &&-{\lambda\over a\rho}\left(\Delta_{\langle a}{\rm curl}\omega_{b\rangle}-a\partial_{\langle a}{\rm curl}\omega_{b\rangle}\right)- \sigma_{c\langle a}\sigma^c{}_{b\rangle}+ \omega_{c\langle a} \omega^c{}_{b\rangle}  \label{isigmadot}
\end{eqnarray}
and
\begin{eqnarray}
\dot{\omega}_{ab}&=& -{2\over3}\,\Theta\omega_{ab}- {2\lambda\over3a^2\rho}\left[\mathcal{Z}_{[a}\Delta_{b]}-{3\over2} \left(a\Delta_{[a}{\rm curl}\omega_{b]} -a^2\partial_{[a}{\rm curl}\omega_{b]}\right)\right] \nonumber\\ &&-2\sigma_{c[a}\omega^c{}_{b]}\,,  \label{iomegadot1}
\end{eqnarray}
respectively. Not surprisingly, we find that viscosity can modify every apsect of the model's kinematics in a variety of ways. Perhaps the most direct effect, relative to the barotropic-fluid case, is seen in Eq.~(\ref{iomegadot1}). The latter shows that viscosity, together with an overall inhomogeneity, can act as a source of rotation (at the second perturbative level).

\subsection{Hydrodynamic flows as purely gravitational
%%%%%%%%%%%%%%%%%%%%%%%%%%%%%%%%%%%%%%%%%%%%%%%%%%%%%%
motions}\label{ssHFPGMs}
%%%%%%%%%%%%%%%%%%%%%%%%
Keplerian motions are central to mass measurements. The observational determination of the masses of various astrophysical systems is usually based on the assumption of purely gravitational motions. For example, the central mass concentration in various galaxies is estimated by Doppler-shift measurements of radiative sources, which are assumed to move along Keplerian trajectories (e.g.~see~\cite{KR}). Nevertheless, there are known cases where the non-gravitational forces are strong enough to affect these trajectories and where a hydrodynamic description of the motion is more appropriate~\cite{HNU}. Then, one would like to know whether the standard measurements have overestimated or underestimated the available amount of matter.

One way of addressing this question is by rewriting key hydrodynamic equations into a ``Keplerian'' form and then examining the implications of such a transformation for the dynamics of the physical system under consideration. Following~\cite{KS}, and in absence of anisotropic pressure, we may combine Eqs.~(\ref{Aa}) and (\ref{CEs}b) to
\begin{equation}
\dot{v}_a= -\partial_a\Phi- {1\over\rho}\,\partial_ap\,.  \label{dotva2}
\end{equation}
Setting $V=1/\rho$ as the specific volume, we introduce an equation of state of the form $E=E(p,V)$, where $E$ is the specific internal energy of the fluid. We may also define the associated temperature $T=T(p,V)$ and specific entropy $S=S(p,V)$ by (e.g.~see~\cite{E1})
\begin{equation}
{\rm d}E+ p\,{\rm d}V= T{\rm d}S\,.  \label{dE}
\end{equation}
where the right-hand side vanishes when adiabaticity holds. In the case of purely isentropic motions (i.e.~when $S$ is spatially and temporally constant), one can use the above expression to recast (\ref{dotva2}) as
\begin{equation}
\dot{v}_a= -\partial_a\Phi- \partial_a\left(E+{p\over\rho}\right)\,, \label{dotva3}
\end{equation}
thus incorporating the internal properties of the fluid into Euler's equation. This means that isentropic hydrodynamic flows can be seen as entirely gravitational motions under the new, effective potential
\begin{equation}
\tilde{\Phi}= \Phi+ E+ {p\over\rho}\,,  \label{tPhi}
\end{equation}
which is shown to correspond to an effective mass-density given by
\begin{equation}
\partial^2\tilde{\Phi}= {1\over2}\,\kappa\tilde{\rho}\,.  \label{trho}
\end{equation}
The ``Keplerial'' density introduced above can be expressed in terms of the wider fluid characteristics, like its internal energy and pressure, through definition (\ref{tPhi}). In general, $\tilde{\rho}$ is different from its hydrodynamic counterpart and their difference
\begin{equation}
{1\over2}\,\kappa(\tilde{\rho}-\rho)= \partial^2\left(E+{p\over\rho}\right)\,,  \label{trho-rho}
\end{equation}
depends on the aforementioned physical properties of the fluid.\footnote{The effective mass density $\tilde{\rho}$ does not generally obey a continuity equation of the simple form (\ref{CEs}a).} This result also offers a way of measuring the ``error-bars'' between mass estimates based on purely gravitational motions, relative to those using the more realistic hydrodynamic approximation. For instance, if the ``virtual'' density $\tilde{\rho}$ is smaller than the ``actual'' one ($\rho$), mass measurements using Keplerian motions will underestimate the available amount of matter. Although the results generally depend on the particulars of the physical system under consideration, there seem to exist realistic astrophysical environments where $\tilde{\rho}<\rho$ (see~\cite{KS,S} for further astrophysical discussion).

\section{Covariant magnetohydrodynamics}\label{sCMHDs}
%%%%%%%%%%%%%%%%%%%%%%%%%%%%%%%%%%%%%%%%%%%%%%%%%%%%%%
Covariant techniques were introduced to the study of electromagnetic fields in~\cite{Eh} and more recently in~\cite{TB} (see also~\cite{BMT} for an up to date review). All these studies are relativistic, however, and so far the Newtonian version of 1+3 covariant electrodynamics and magnetohydrodynamics (MHD) has been missing from the literature.

\subsection{Maxwell's equations}\label{ssMEs}
%%%%%%%%%%%%%%%%%%%%%%%%%%%%%%%%%%%%%%%%%%%%%
In a two-fluid plasma description the charge carriers are the positive ions and the electrons, which are treated as two coupled conducting fluids. The matter density, the charge density and the current density of the one-fluid description are
\begin{equation}
\rho= m_+n_++ m_-n_-\,, \hspace{15mm} q= e(n_+-n_-)  \label{rho-q}
\end{equation}
and
\begin{equation}
{\cal J}_a= e(n_+v^+_a-n_-v^-_a)\,,  \label{cJ}
\end{equation}
respectively (e.g.~see~\cite{G}). In the above $e$ is the electron charge, $m_{\pm}$, are the ion and the electron masses, $n_{\pm}$ represent their
number densities and $v_a^{\pm}$ are the associated velocities. In the case of global electric neutrality, we have $n_+=n_-$ and the centre of mass of the ion-electron system has the ``bulk'' velocity~\footnote{See~\cite{CS} for a generalisation to general relativity and further discussion.}
\begin{equation}
v_a={1\over m_++m_-}\,(m_+v_a^++m_-v_a^-)\,.  \label{cmvel}
\end{equation}

Within the single fluid approach and at the limit of resistive magnetohydrodynamics (MHD), the displacement current ($\partial_tE_a$) is negligible. Then, Maxwell's equations reduce into a set of one propagation equation
\begin{equation}
\partial_tB_a= -{\rm curl}E_a\,,  \label{M1}
\end{equation}
and three constraints
\begin{equation}
{\rm curl}B_a= {\cal J}_a\,, \hspace{20mm} \partial^aE_a=0
\hspace{15mm} {\rm and} \hspace{15mm} \partial^aB_a=0\,,  \label{M234}
\end{equation}
having adopted the Heaviside-Lorentz electromagnetic units. The above, which respectively correspond to Faraday's law, Amp\`ere's law, Coulomb's law and Gauss' law, are  supplemented by Ohm's law. For a fluid with nonzero electrical resistivity, the latter reads
\begin{equation}
{\cal J}_a=\varsigma(E_a+\epsilon_{abc}v^bB^c)\,,  \label{rOhm}
\end{equation}
with $\varsigma$ representing the (scalar) electrical conductivity of the medium. This form of Ohm's law corresponds to the resistive MHD approximation, which applies to fluids with small but finite electrical resistivity. In general, Eq.~(\ref{rOhm}) contains several additional terms -- like those representing the Hall and the Biermann-battery effects (e.g.~see expression (3.5.9) in~\cite{KT}).\footnote{The potential implications of a non-conventional (anomalous) form of electrical resistivity were discussed in~\cite{VTP}. Also, for a comparison with the fully relativistic counterpart of (\ref{rOhm}), the reader is referred to~\cite{KaT}.}

Solving (\ref{rOhm}) for the electric field vector, substituting the result into Eqs.~(\ref{M1}), using decomposition (\ref{pbva}), constraint (\ref{M234}a) and involving the convective derivative operator, we obtain the covariant form of the Newtonian magnetic induction equation
\begin{equation}
\dot{B}_a= -{2\over3}\,\Theta B_a+ (\sigma_{ab}+\omega_{ab})B^b+ {1\over\varsigma}\,\partial^2B_a\,.  \label{rM1}
\end{equation}
at the resistive MHD limit. Comparing the above to its relativistic counterpart (see Eq.~(3.2.4) in~\cite{BMT}) we notice that the relative motion terms (i.e.~the first two terms in the right-hand side of the two formulae) are identical. We also note the absence of the acceleration terms from (\ref{M234}a) and (\ref{rM1}) -- compare the former to (3.2.3) in~\cite{BMT}. This absence reflects the fact that Newtonian physics treats time and space as entirely separate entities.

Similarly, employing (\ref{rOhm}) together with Amp\`ere's law (see Eq.~(\ref{M234}a)), expression (\ref{M234}b) -- Coulomb's law -- assumes the covariant form
\begin{equation}
2B^a\omega_a= -v^a{\cal J}_a\,,  \label{rM3}
\end{equation}
suggesting that the sum $v^a\mathcal{J}_a=v^a{\rm curl}B_a$ acts as an effective charge density relative to a rotating observer (when $B^a\omega_a\neq0$ -- see also Eq.~(3.2.5) in~\cite{BMT}).

\subsection{Magnetic evolution}\label{ssME}
%%%%%%%%%%%%%%%%%%%%%%%%%%%%%%%%%%%%%%%%%%%
Contracting the magnetic induction equation (see (\ref{rM1})) along the field vector leads to the nonlinear evolution formula of the magnetic pressure, namely to
\begin{equation}
\left(B^2\right)^{\cdot}= -{4\over3}\,\Theta B^2- 2\sigma^{ab}\Pi_{ab}+ {1\over\varsigma}\, \partial^2B^2- {2\over\varsigma}\,\left[\left(\partial_{\langle b} B_{a\rangle}\right)^2-\left(\partial_{[b} B_{a]}\right)^2\right]\,,  \label{rmceq1}
\end{equation}
where
\begin{equation}
\Pi_{ab}= -B_{\langle a}B_{b\rangle}= {1\over3}\,B^2\delta_{ab}- B_aB_b\,,  \label{Pi}
\end{equation}
by definition. The latter is a symmetric, trace-free tensor that describes the magnetic anisotropic stresses and corresponds precisely to its relativistic counterpart. By definition, $\Pi_{ab}=\mathcal{M}_{ab}- (B^2/6)\delta_{ab}$, where $\mathcal{M}_{ab}=(B^2/2)\delta_{ab}-B_aB_b$ is the Maxwell tensor~\cite{P}. Thus, in agreement with the relativistic analysis (see section \S~5.1 in~\cite{BMT}), the $B$-field exerts an isotropic pressure equal to $p_B=\mathcal{M}_a{}^a/3= B^2/6$ and has an anisotropic pressure component given by $\Pi_{ab}$. We also note the quantities $\partial_{\langle b} B_{a\rangle}$ and $\partial_{[b} B_{a]}$ in the right-hand side of (\ref{rmceq1}). These may be respectively interpreted a the shear and the vorticity analogues of the $B$-field -- see Eq.~(\ref{MHDRay}) below -- and are important in highly distorted and turbulent magnetic configurations.

Definition (\ref{Pi}) immediately ensures that $\Pi_{ab}B^b=-(2B^2/3)B_a$. This in turn means that the $B$-field is an eigenvector of the $\Pi_{ab}$-tensor, with $-2B^2/3$ being the associated eigenvalue. The negative sign shows that the magnetic pressure in the direction of the field lines is negative and reflects the tension properties of the latter (see also \S~\ref{ssLF} below). Projecting (\ref{Pi}) orthogonal to the magnetic forcelines, on the other hand, we find a positive eigenvalue equal to $B^2/3$, which verifies that the field exerts a positive pressure in that plane~\cite{TM2}. In other words, every single field line acts like an elastic rubber band under tension, while neighbouring lines tend to push each other apart~\cite{P}.

Finally, following (\ref{Pi}), it becomes immediately clear that the magnetic induction equation -- together with expression (\ref{rmceq1}) -- also monitors the time evolution of the anisotropic pressure of the field. On the other hand, the divergence of (\ref{Pi}) provides the associated constraint, namely
\begin{equation}
\partial^b\Pi_{ab}= \varepsilon_{abc}B^b{\rm curl}B^c- {1\over6}\,\partial_aB^2\,.
\label{divPi}
\end{equation}

\section{Resistive magnetohydrodynamics}\label{sRMHD}
%%%%%%%%%%%%%%%%%%%%%%%%%%%%%%%%%%%%%%%%%%%%%%%%%%%%%
The resistive (or real) MHD scheme is believed to provide a good approximation to a variety of ``typical'' physical environments. For example, when the Larmor frequency and bulk velocity of the plasma are small, or when the dimensions of the system under study are large.

\subsection{The Lorentz force}\label{ssLF}
%%%%%%%%%%%%%%%%%%%%%%%%%%%%%%%%%%%%%%%%%%
In covariant terms, the evolution of a non-relativistic magnetised plasma of finite electrical resistivity is monitored by the nonlinear set
\begin{equation}
\dot{\rho}= -\Theta\rho\,, \label{rMHDCe}
\end{equation}
\begin{equation}
\rho A_a= -\partial_ap- \partial^b\pi_{ab}- \varepsilon_{abc}B^b {\rm curl}B^c\,,  \label{rMHDEe}
\end{equation}
\begin{equation}
\partial^2\Phi= {1\over2}\,\kappa\rho\,,  \label{rMHDPe}
\end{equation}
consisting of the continuity equation, the Navier-Stokes equation and Poisson's formula respectively. These are supplemented by Maxwell's equations, which applied to an electrically resistive medium and written in covariant form read
\begin{equation}
\dot{B}_a= -{2\over3}\,\Theta B_a+ \left(\sigma_{ab}+\omega_{ab}\right)B^b+ {1\over\varsigma}\,\partial^2B_a\,,  \label{rMHDM1}
\end{equation}
\begin{equation}
{\rm curl B}_a=\mathcal{J}_a\,,  \label{rMHDM2}
\end{equation}
\begin{equation}
\partial^aB_a= 0\,,  \label{rMHDM3}
\end{equation}
and
\begin{equation}
2B^a\omega_a= -v^a\mathcal{J}_a= -v^a{\rm curl}B_a\,.  \label{rMHDM4}
\end{equation}
The momentum conservation is reflected in (\ref{rMHDEe}), which is the Navier-Stokes equation generalised to a (globally neutral) magnetised fluid. This expression can be obtained directly from its hydrodynamic counterpart (see (\ref{CEs}b)) by implementing the aforementioned fluid description of the $B$-field. To be precise, Eq.~(\ref{rMHDEe}) emerges after replacing $p$ with $p+B^2/6$ and $\pi_{ab}$ with $\pi_{ab}+\Pi_{ab}$ in the right-hand side of (\ref{CEs}b), while using constraint (\ref{divPi}) at the same time. Note that there is no magnetic contribution to the total inertial mass in the left-hand side of (\ref{rMHDEe}), or to the total gravitational mass in the right-hand side of Eq.~(\ref{MHDRay}) -- see \S~\ref{ssNKs} below. This is a significant change, with respect to the relativistic case (compare to expressions (5.3.3) and (5.5.1) of~\cite{BMT}), which implies that there is no Newtonian analogue to the relativistic energy density of the $B$-field. Finally, following (\ref{rMHDEe}), we note that the sum $A_aB^a$ has no magnetic dependence. This ensures that the magnetic field has no effect along its own direction.

The set (\ref{rMHDCe})-(\ref{rMHDM4}) is supplemented by the kinematic propagation and constraint equations (\ref{Ray})-(\ref{kcon3}), once the latter have been appropriately adapted to our electrically resistive magnetised environment. Within the limits of the Newtonian theory, the above named formulae contain no explicit magnetic terms. This means that the kinematic effects of the $B$-field propagate solely through the fluid acceleration and specifically via the Lorentz-force term in the right-hand side of Eq.~(\ref{rMHDEe}).\footnote{The lack of explicit magnetic terms in the propagation formulae (\ref{Ray})-(\ref{omegadot2}) and the absence of acceleration terms in Eqs.~(\ref{kcon1})-(\ref{kcon3}), represents a considerable change relative to the relativistic case (see~\cite{BMT} for details).} For a globally neutral medium, the Lorentz force depends exclusively on the $B$-field and splits into two stresses according to
\begin{equation}
\varepsilon_{abc}B^b{\rm curl}B^c= {1\over2}\,\partial_aB^2- B^b\partial_bB_a\,,  \label{Lorentz1}
\end{equation}
where the first term in the right-hand side is due to the isotropic pressure of the field (see \S~\ref{ssME} below) and the second carries the effects of the magnetic tension. The tension stress also reflects the elasticity of the field lines and their tendency to remain straight~\cite{P}. When these two stresses balance each other out, the $B$-field reaches equilibrium.

\subsection{Nolinear kinematics}\label{ssNKs}
%%%%%%%%%%%%%%%%%%%%%%%%%%%%%%%%%%%%%%%%%%%%%
Proceeding as in \S~\ref{ssPFs}, we ignore the anisotropic pressure of the fluid and assume a barotropic medium by setting $p=p(\rho)$. Then, the MHD version of the Navier-Stokes equation (see (\ref{rMHDEe})) takes the form
\begin{equation}
A_a= -{{\rm c}_s^2\over a}\,\Delta_a- {{\rm c}_a^2\over2a}\,\mathcal{B}_a+ {1\over\rho}\,B^b\partial_bB_a\,,  \label{MHDNS}
\end{equation}
with ${\rm c}_a^2=B^2/\rho$ and $\mathcal{B}_a=(a/B^2)\partial_aB^2$. The former is the Alfv\'en speed, which determines the propagation of MHD disturbances and also provides a measure of the relative strength of the $B$-field. The latter is a dimensionless variable that describes spatial variations in the (isotropic) magnetic pressure. Note the last term in the right-hand side of Eq.~(\ref{MHDNS}), which carries the effects of the magnetic tension (see decomposition (\ref{Lorentz1}) above). When the sound speed, the Alfv\'en speed and the scale-factor have a spatially homogeneous distribution, the gradient of the above leads to
\begin{equation}
\partial_bA_a= -{{\rm c}_s^2\over a^2}\,\Delta_{ab}- {{\rm c}_a^2\over2a^2}\,\mathcal{B}_{ab}- {1\over a\rho}\, \Delta_bB^c\partial_cB_a+ {1\over\rho}\,\partial_bB^c\partial_cB_a+ {1\over\rho}\,B^c\partial_c\partial_bB_a\,,  \label{MHDpbAa}
\end{equation}
where $\mathcal{B}_{ab}=a\partial_b\mathcal{B}_a$. The overall magnetic effect is rather involved and propagates via the last four terms. Of these, the first is triggered by the isotropic pressure of the field and the rest are due to the tension properties of the magnetic forcelines.

Substituting the trace of (\ref{MHDpbAa}) into Eq.~(\ref{Ray}), we obtain the nonlinear form of Raychauduri's formula for a magnetised, self-gravitating Newtonian fluid of zero total charge. In particular, using constraint (\ref{M234}c), we arrive at
\begin{eqnarray}
\dot{\Theta}&=& -{1\over3}\,\Theta^2- {1\over2}\,\kappa\rho- {{\rm c}_s^2\over a^2}\,\Delta- {{\rm c}_a^2\over2a^2}\,\mathcal{B}- {1\over a\rho}\,\Delta^aB^b\partial_bB_a- 2\left(\sigma^2-\sigma_B^2\right) \nonumber\\ &&+2\left(\omega^2-\omega_B^2\right)\,,  \label{MHDRay}
\end{eqnarray}
where $\mathcal{B}=\mathcal{B}^a{}_a$, $\sigma_B^2= \partial_{\langle b}B_{a\rangle}\partial^{\langle b}B^{a\rangle} /2\rho$ and $\omega_B^2=\partial_{[b}B_{a]}\partial^{[b}B^{a]} /2\rho$. The former describes scalar variations in the magnetic pressure, while the last two can be interpreted as the magnetic analogues of the shear and the vorticity respectively. According to the above, the compression of the field lines (which corresponds to an increase in the magnetic pressure and $\mathcal{B}>0$) assists the gravitational pull of the matter. The dilution of the magnetic forcelines, on the other hand, acts against contraction. We also note that the effect of the magnetic shear and vorticity opposes that of their kinematic counterparts (see also~\cite{PE}). The reason behind this counterintuitive behaviour is the magnetic tension. Both $\sigma_B$ and $\omega_B$ are triggered by the elasticity of the magnetic forcelines and therefore react to any agent that  distorts them. The magnetic vorticity, in particular, is the response of the field's tension to the twisting of its forcelines. The resulting stress slows the rotation down and this effect is commonly referred to as ``magnetic braking''. Analogous behaviour has also been observed in relativistic studies (see~\cite{T} for more details and further discussion). The key difference here, as a result of the Euclidean nature of the Newtonian space, is the absence of the general relativistic magneto-curvature stresses.

Substituting the symmetric and trace-free component of the auxiliary expression (\ref{MHDpbAa}) into the right-hand side of (\ref{sigmadot}), leads to
\begin{eqnarray}
\dot{\sigma}_{ab}&=& -{2\over3}\,\Theta\sigma_{ab}- E_{ab}- {{\rm c}_s^2\over a^2}\,\Delta_{\langle ab\rangle}- {{\rm c}_a^2\over2a^2}\,\mathcal{B}_{\langle ab\rangle}+ {1\over\rho}\,B^c\partial_c\partial_{\langle b}B_{a\rangle}+ {1\over\rho}\,\partial_{\langle b}B^c\partial_cB_{a\rangle} \nonumber\\ &&-{1\over a\rho}\,B_c\Delta_{\langle a}\partial^cB_{b\rangle}- \sigma_{c\langle a}\sigma^c{}_{b\rangle}+ \omega_{c\langle a} \omega^c{}_{b\rangle}\,.  \label{MHDsigmadot}
\end{eqnarray}
The above shows how anisotropies in the distribution of the magnetic pressure and in that of the field gradients affect the evolution of the kinematic shear. In particular, despite the lack of a direct contribution from the magnetic anisotropic pressure, the $B$-field acts as a shear source in a variety of ways.\footnote{Recall that in relativistic studies the magnetic $\Pi_{ab}$-tensor is an explicit source of shear anisotropies~\cite{BMT}.} Note that of the four magnetic source-terms in Eq.~(\ref{MHDsigmadot}), the first is due to the field's pressure and the last three are the result of its tension.

Finally, the skew part of decomposition (\ref{MHDpbAa}), together with the (strictly Newtonian) results $\mathcal{B}_{[ab]}=0=\Delta_{[ab]}$, transform Eq.~(\ref{omegadot2}) into
\begin{equation}
\dot{\omega}_a= -{2\over3}\,\Theta\omega_a- {1\over2\rho}\, B^b\partial_b{\rm curl}B_a- {1\over2\rho}\, \varepsilon_{abc}\partial^bB^d\partial_dB^c+ {1\over2a\rho}\, \varepsilon_{abc}B_d\Delta^b\partial^dB^c+ \sigma_{ab}\omega^b\,. \label{MHDomegadot}
\end{equation}
This expression reveals the role of the $B$-field as a source of rotation, either on its own or through its coupling to the density gradients. It should also be noted that there are no effects due to the isotropic magnetic pressure in Eq.~(\ref{MHDomegadot}), with all the $B$-terms coming from the field's tension. Following (\ref{MHDsigmadot}) and (\ref{MHDomegadot}), even if the fluid is originally shear-free and non-rotating, it will not remain so once a magnetic field is introduced.

We close this section by noting that, according to Eqs.~(\ref{kcon1})-(\ref{kcon3}), the kinematic constraints contain no explicit magnetic terms and therefore are only indirectly affected by the field's presence. We should also underline the benefits from using the covariant approach. These are multiple because the formalism streamlines the equations, while maintaining maximum detail and physical transparency. Finally, we note that the expressions given in \S~\ref{sRMHD} can be used to study the Newtonian evolution of any electrically resistive and globally neutral fluid in the presence of a magnetic field. In addition, the formalism developed so far can be extended to study the behaviour of inhomogeneities, both at the linear and at the nonlinear level.

\subsection{Hydrodynamic reduction of magnetised flows}
%%%%%%%%%%%%%%%%%%%%%%%%%%%%%%%%%%%%%%%%%%%%%%%%%%%%%%%
Ideal fluids have zero anisotropic pressure by definition. When, in addition, the tension component of the magnetic Lorentz force is also zero (i.e.~for $B^b\partial_bB_a=0$ -- see decomposition (\ref{Lorentz1})), the generalised Navier-Stokes equation simplifies to
\begin{equation}
\rho A_a= -\partial_ap- {1\over2}\,\partial_aB^2\,.  \label{NS2}
\end{equation}
Realistically speaking, the above is only an approximation and holds when the Lorentz force is dominated by the (positive) pressure of the $B$-field. In such an environment, the non-gravitational acceleration of the fluid (i.e.~the vector $A_a$) comes purely from a potential. Then, the MHD motion reduces to a simple hydrodynamic flow with
\begin{equation}
\rho A_a= -\partial_aP\,,  \label{eE}
\end{equation}
where the scalar $P=p+B^2/2$ acts as an effective hydrodynamic pressure (see Eq.~(\ref{MHDNS})). This new motion is monitored by the formulae of \S~\ref{ssNHD}, after replacing expression (\ref{CEs}b) with (\ref{eE}) and the pressure of the original fluid with the above given effective pressure $P$. One can also go a step further and use the transformations of \S~\ref{ssHFPGMs} to represent the MHD flow of (\ref{NS2}), (\ref{eE}) as a ``purely gravitational'' motion. This time the effective potential will also depend on the magnetic pressure.

\section{Ideal magnetohydrodynamics}\label{sIMHD}
%%%%%%%%%%%%%%%%%%%%%%%%%%%%%%%%%%%%%%%%%%%%%%%%%
In most astrophysical and cosmological studies, magnetic fields are treated within the limits of the ideal MHD approximation. The latter, applies to highly conductive media with essentially zero electrical resistivity. Although overly idealised and simplistic, the perfect MHD scheme still seems to provide the correct description in a variety of studies.

\subsection{Maxwell's equations}\label{ssIMEs}
%%%%%%%%%%%%%%%%%%%%%%%%%%%%%%%%%%%%%%%%%%%%%%
When dealing with a perfectly conductive medium, namely at the $\varsigma\rightarrow\infty$ limit, the Ohmic current in Eq.~(\ref{rOhm}) vanishes (i.e.~$\mathcal{J}_a/\varsigma\rightarrow0$) and the associated electric field is given by the simple expression
\begin{equation}
E_a=-\epsilon_{abc}v^bB^c\,.  \label{icOhm}
\end{equation}
In these environments the nonlinear equations monitoring a globally neutral, self-gravitating Newtonian fluid in the presence of a magnetic field are
\begin{equation}
\dot{\rho}= -\Theta\rho\,, \label{MHDCe}
\end{equation}
\begin{equation}
\rho A_a= -\partial_ap- \partial^b\pi_{ab}- \varepsilon_{abc}B^b {\rm curl}B^c\,,  \label{MHDEe}
\end{equation}
\begin{equation}
\partial^2\Phi= {1\over2}\,\kappa\rho\,,  \label{MHDPe}
\end{equation}
\begin{equation}
\dot{B}_a= -{2\over3}\,\Theta B_a+ \left(\sigma_{ab}+\omega_{ab}\right)B^b\,,  \label{MHDM1}
\end{equation}
\begin{equation}
{\rm curl B}_a= \mathcal{J}_a\,,  \label{MHDM2}
\end{equation}
\begin{equation}
\partial^aB_a= 0\,,  \label{MHDM3}
\end{equation}
\begin{equation}
2B^a\omega_a= -v^a\mathcal{J}_a= -v^a{\rm curl}B_a\,.  \label{MHDM4}
\end{equation}
Relative to the resistive-MHD case of the section \S~\ref{sRMHD}, we note the absence of a diffusion term in the right-hand side of the induction equation (compare Eqs.~(\ref{rM1}), (\ref{rMHDM1}) to expression (\ref{MHDM1}) above).\footnote{The kinematics of a perfectly conductive ideal fluid are still monitored by the ``resistive'' formulae of \S~\ref{ssNKs}.} This guarantees that the magnetic field lines remain frozen-in with the matter. In particular, (\ref{MHDM1}) ensures that $\mathcal{X}_a=a^3B_a$ is a relative position vector connecting the same particles at all times (i.e.~$\dot\mathcal{X}_a=\mathcal{X}^b\partial_bv_a$ -- see footnote~\ref{rvv} in \S~\ref{ssS-GFs} and also~\cite{E2,BMT}).

\subsection{Magnetic evolution}\label{ssIME}
%%%%%%%%%%%%%%%%%%%%%%%%%%%%%%%%%%%%%%%%%%%%
Relation (\ref{MHDM1}) also shows that, in the absence of shear anisotropies, the magnetic strength either dilutes with the expansion or increases with the contraction of the fluid. Then, recalling that $\Theta/3=\dot{a}/a$, the magnetic induction equation reduces to
\begin{equation}
\dot{B}_a= -2\left({\dot{a}\over a}\right)B_a\,.  \label{MHDie}
\end{equation}
An immediate consequence ia that the magnetic flux, here represented by the quantity $a^2B_a$, remains conserved in time. Moreover, the ideal-MHD counterpart of Eq.~(\ref{rmceq1}) reads
\begin{equation}
\left(B^2\right)^{\cdot}= -{4\over3}\,\Theta B^2- 2\sigma^{ab}\Pi_{ab}\,,  \label{rmceq2}
\end{equation}
with $\Pi_{ab}$ given in (\ref{Pi}). Therefore, for zero shear anisotropy, we recover the familiar from the relativistic studies radiation-like evolution (i.e.~$B^2\propto a^{-4}$) of the magnetic pressure. The presence of shear, on the other hand, will generally modify the aforementioned ``adiabatic'' pattern. This can happen during the realistic (i.e.~anisotropic) collapse of a magnetised proto-galactic cloud and lead to the amplification of the embedded $B$-field beyond the limits of the simple spherical-collapse models~\cite{DBL}.

As mentioned in \S~\ref{ssME}, the magnetic induction equation monitors the time evolution of both the isotropic and the anisotropic pressure of the $B$-field. At the ideal-MHD limit the time derivative of (\ref{Pi}) combines with expressions (\ref{MHDie}) and (\ref{rmceq2}) to give
\begin{equation}
\dot{\Pi}_{ab}= -{4\over3}\,\Theta\Pi_{ab}+ 2\Pi_{c\langle
a}\sigma^c{}_{b\rangle}- 2\Pi_{c\langle a}\omega^a{}_{b\rangle}-
{2\over3}\,B^2\sigma_{ab}\,,  \label{dotPi}
\end{equation}
while the associated constraint is still given by (\ref{divPi}). In the absence of shear and vorticity, the above leads to $\Pi_{ab}\propto a^{-4}$, in line with the evolution of its isotropic counterpart. Thus, when the anisotropy is small, the $B$-field has a radiation-like evolution to first approximation.

Turning to the kinematics of perfectly conducting media, we note that the magnetic effects on a (globally neutral) fluid propagate via the Lorentz-force term in the right-hand side of the generalised Navier-Stokes formula (see Eq.~(\ref{MHDEe})). The form of the latter is independent of the electrical resistivity of the matter, since it contains no related terms. This means that relations (\ref{MHDRay})-(\ref{MHDomegadot}), together with constraints (\ref{kcon1})-(\ref{kcon3}), also govern the kinematics of an ideal-MHD medium.\footnote{Directly, the electrical resistivity of the medium affects only the evolution of the embedded $B$-field (compare Eqs.~(\ref{rMHDM1}) and (\ref{MHDM1})). The latter then carries these effects to the kinematics and the dynamics of the magnetised medium.} When an equation of the state for the matter is introduced, these expressions monitor the nonlinear evolution of the magnetised medium completely and in a fully covariant manner.

We finally note that, when the fluid is perfect, the magnetic field becomes the sole source of anisotropy. The magnetically induced effective viscosity can be related to that of the shear in a way that closely resembles the phenomenological equation of state introduced in \S~\ref{ssIFs} (see Eq.~(\ref{pi}) there). Thus, assuming that the $B$-field is a shear eigenvector, we may set $\sigma_{ab}B^b=(2\mu/3)B_a$, where $2\mu/3$ is the associated eigenvalue. Also, following definition (\ref{Pi}), we find that $\Pi_{ab}B^b=-(2B^2/3)B_a$ and subsequently arrive at
\begin{equation}
\Pi_{ab}= -\lambda\sigma_{ab}\,,  \label{Pi-sigma}
\end{equation}
with $\lambda=B^2/\mu$ acting as an effective coefficient of magnetic viscosity~\cite{TM2}.

\section{Discussion}\label{sD}
%%%%%%%%%%%%%%%%%%%%%%%%%%%%%%
Newtonian theory offers a very good approximation to general relativity in weak-gravity environments and also on scales well inside the Hubble radius. The covariant approach to Newtonian hydrodynamics is a Lagrangian description based on a relative-motion treatment that exploits the irreducible kinematical quantities of the motion. Although the formalism was originally applied within the framework of Newton's theory, it has since been used primarily in relativistic studies. On the other hand, while the 1+3-covariant techniques have been employed for the study of relativistic electromagnetic fields, so far a Newtonian version of that work has been missing.

The present paper reviews and extends the existing work on Newtonian covariant hydrodynamics on the one hand, while on the other it applies the covariant techniques to magnetohydrodynamic studies. Exploiting the advantages of the relative-motion treatment, we supplement the standard hydrodynamic formulae with a set of three propagation and three constraint equations that monitor the evolution of the irreducible kinematical variables. The latter, namely the volume expansion/contraction, the shear and the vorticity, describe the relative motion of neighbouring flow lines. The aforementioned formulae are obtained in a manner analogous to that of their relativistic counterparts and this facilitates the direct comparison of the two sets. In fact, the close analogy between the Newtonian and the relativistic equations is maintained throughout the paper and this allows the unambiguous identification of their differences.

We consider perfect as well as imperfect (viscous) media, looking for differences in their kinematical behaviour. Not surprisingly, the extra degree of freedom that viscosity introduces, means that the kinematics of a viscous fluid are considerably more involved. Following our analysis, the key contribution of  viscosity is perhaps through its role as a source of rotation. Focusing on isentropic fluids, we also discuss how hydrodynamic flows can be represented as ``purely gravitational'' motions due to a new (effective) potential. The latter incorporates additional characteristics of the fluid, like its internal energy and pressure, and corresponds to a new (effective) mass density. The relation between the ``actual'' (the hydrodynamic) and the ``virtual'' (the effective) mass density has been used to estimate the accuracy of astrophysical mass measurements based on the assumption of purely gravitational (Keplerian) motions.

Assuming an imperfect MHD fluid of zero total charge, we derive the covariant version of Maxwell's equations within the limits of Newtonian gravity. In an environment of small but finite electrical resistivity, we monitor the evolution of the magnetic field completely. This means providing the nonlinear propagation and constraint equations for both the isotropic and the anisotropic magnetic pressure. The compactness of the covariant formalism also allows us to identify the impact of the $B$-field on the kinematics of the fluid in detail. In practice, this means isolating the effects due to the ordinary (the positive) magnetic pressure, from those coming from the tension of its forcelines. The former affect the volume evolution and also the shape of a given fluid element, but not its rotational behaviour. The impact of the magnetic tension, on the other hand, is more widespread and sometimes counterintuitive. Thus, the elastic properties of the field lines are shown to act as sources of rotation, either on their own or through their coupling to density inhomogeneities. In an analogous way, magnetism is also found to trigger shear distortions. Moreover, when looking into the implications of the $B$-field for the volume evolution of the fluid, we identify magnetic analogues of the shear and the vorticity. Both carry the tension properties of the field and oppose the effects of their kinematic counterparts. The ``magnetic vorticity'' term, in particular, tends to slow the rotation down and leads to what is commonly referred to as magnetic braking. Finally, we apply our analysis to the limit of ideal magnetohydrodynamics and also discuss how certain MHD flows can be reduced to simple hydrodynamic ones.

The formalism developed here can be applied to a variety of astrophysical and cosmological environments, where the Newtonian theory is a good approximation. This includes non-relativistic astrophysical MHD and galaxy formation studies. In the latter case, for example, one could use the linearised version of our equations to follow the linear regime of a magnetised protogalactic cloud (with size well below the horizon scale). Similarly, the full expressions can be employed to monitor the nonlinear evolution of the protogalaxy, once the latter has decoupled from the background expansion and started collapsing. More specifically, our equations will enable one to look for effects outside the limits of the ideal MHD. The latter are expected to play a role during the nonlinear regime of galaxy formation (at least locally). Applications of this type will be the subject of future work.

\end{document}